\documentstyle[12pt]{article}
\input epsf
\newcommand{\sect}[1]{\section{#1}\setcounter{equation}{0}}

\renewcommand{\*}{ \hspace{-6pt}&=&\hspace{-6pt} }

\newcommand{\ol}[1]{ \hspace{1pt}\overline{\hspace{-1pt}#1
   \hspace{-1pt}}\hspace{1pt} }
\def\IP{{\rm I\!P}}
\begin{document}

\newpage
\bigskip
\hskip 3.7in\vbox{\baselineskip12pt
\hbox{NSF-ITP-99-122}
\hbox{IASSNS-HEP-99/107}
\hbox{DTP/99/81}
\hbox{hep-th/9911161}}

\bigskip\bigskip

\centerline{\large \bf  Gauge Theory and the Excision
 of Repulson Singularities}

\bigskip\bigskip

\centerline{{\bf
Clifford V. Johnson\footnote{c.v.johnson@durham.ac.uk},
Amanda W. Peet\footnote{peet@itp.ucsb.edu},
Joseph Polchinski\footnote{joep@itp.ucsb.edu}}}
\bigskip
\smallskip
\centerline{$^1$School of Natural Sciences}
\centerline{Institute for Advanced Study}
\centerline{Princeton, NJ 08540, U.S.A.}

\bigskip
\centerline{$^1$Centre For Particle Theory}
\centerline{Department of Mathematical Sciences}
\centerline{University of Durham, Durham DH1 3LE, U.K.}

\bigskip
\centerline{$^{2,3}$Institute for Theoretical Physics}
\centerline{University of California}
\centerline{Santa Barbara, CA\ \ 93106-4030, U.S.A.}

\bigskip

\begin{abstract}
\baselineskip=16pt
We study brane configurations that give rise to large-$N$ gauge
theories with eight supersymmetries and no hypermultiplets.  These
configurations include a variety of wrapped, fractional, and stretched
branes or strings.  The corresponding spacetime geometries which we
study have a distinct kind of singularity known as a repulson.  We
find that this singularity is removed by a distinctive mechanism,
leaving a smooth geometry with a core having an enhanced gauge
symmetry.  The spacetime geometry can be related to large-$N$
Seiberg--Witten theory.
\end{abstract}
\newpage
\baselineskip=18pt
\setcounter{footnote}{0}

%--------------------------------------------------------------------+
\sect{Introduction}

Understanding the physics of spacetime singularities is a challenge
for any complete theory of quantum gravity.  It has been shown that
string theory resolves certain seeming singularities, such as
orbifolds~\cite{orbif}, flops~\cite{flop}, and conifolds~\cite{conif},
in the sense that their physics is completely nonsingular.  On the
other hand, it has also been argued that certain singularities should
not be resolved, but rather must be disallowed configurations --- in
particular, negative mass Schwarzschild, which would correspond to an
instability of the vacuum~\cite{value}.  Also, in the study of
perturbations of the AdS/CFT duality various singular spacetimes have
been encountered, and at least some of these must be unphysical in the
same sense as negative mass Schwarzschild.  A more general
understanding of singularities in string theory is thus an important
goal.

In this paper we study a naked singularity of a particular
type~\cite{rep1,rep2,rep3}, which has been dubbed the {\it repulson}.
A variety of brane configurations in string theory appear to give rise
to such a singularity.  However, we will argue that this is not the
case.  Rather, as the name might suggest, the constituent branes
effectively repel one another (in spite of supersymmetry), forming in
the end a nonsingular shell.

Our interest in this singularity arose from a search for new examples
of gauge/gravity duality.  In particular, the brane configurations
that give rise to the repulson singularity have on their world-volumes
pure $D=4$, ${\cal{N}}=2$ gauge theory (or the equivalent in other
dimensions), as opposed to the usual pure $D=4$, ${\cal{N}}=4$, or
$D=4$, ${\cal{N}}=2$ with hypermultiplets.  We do not precisely find
such a duality, in the sense of using supergravity to calculate
properties of the strongly coupled gauge theory, but we do find a
striking parallel between the moduli space of the large-$N$ $SU(N)$
gauge theory and the fate that we have deduced for the singularity.
We also find some clues which allow us to guess at aspects of a
possible dual.

In section~2 we describe the repulson singularity and the various
brane configurations where it arises.  In section~3 we deduce the
specific physical mechanism by which it is removed.  In section~4 we
relate this to behavior of Yang--Mills theory with eight
supersymmetries.  We do not find a duality, in the sense of being able
to use supergravity to calculate in the strongly coupled gauge theory,
but we find a striking parallel between the physical picture deduced
in section~3 and the large-$N$ Seiberg--Witten theory.  We point out
some features suggestive of a dual theory, and remark upon the case
of finite temperature.  In section~5 we develop
two of the dual versions, in terms of bent NS5-branes, and
wrapped/fractional D-branes.  Section~6 offers brief conclusions,
and suggestions for future directions.

%--------------------------------------------------------------------+
\sect{The Repulson Singularity}

Let us consider first the oft-discussed D1--D5 system,
\begin{eqnarray}
ds^2 \* Z_1^{-1/2} Z_5^{-1/2} \eta_{\mu\nu} dx^\mu dx^\nu + Z_1^{1/2}
Z_5^{1/2} dx^i dx^i + Z_1^{1/2} Z_5^{-1/2} dx^m dx^m \ ,\nonumber
\\
e^{2\Phi } \* g^2 {Z_1}/{ Z_5}\ , \nonumber\\
C_{\it 2} \* ({Z_1} g)^{-1} dx^0 \wedge dx^5\ , \nonumber\\
C_{\it 6} \* ({Z_5} g)^{-1} dx^0 \wedge dx^5 \wedge
dx^6 \wedge dx^7 \wedge dx^8 \wedge dx^9
\ . \label{15sol}
\end{eqnarray}
Here $\mu,\nu$ run over the 05-directions tangent to all the branes, $i$
runs over the 1234-directions transverse to all branes, and $m$ runs over
the 6789-directions of a $T^4$, tangent to the D5-branes and transverse to
the D1-branes.  We have defined
\begin{eqnarray}
Z_1 \* 1+\frac{r_1^2 }{ r^2}\ ,\quad  r_1^2 =
\frac{ (2\pi)^4 g Q_1 \alpha'^3 }{ V } \ , \nonumber\\
Z_5 \* 1+\frac{r_5^2 }{ r^2}\ ,\quad  r_5^2 = gQ_5\alpha' \ ,
\end{eqnarray}
with $r^2 = x^i x^i$ and $V$ the volume of the $T^4$.

This configuration
leaves 8 unbroken supersymmetries, all of which transform as $({\bf 2},{\bf
1})$ under the $SO(4)$ that acts on $x^i$.  At the horizon, $r \to 0$, the
geometry approaches AdS$_3 \times S^3 \times T^4$, giving rise to an
AdS/CFT duality~\cite{juan}.
The tension of the effective string in 6 dimensions is
\begin{equation}
\tau = \frac{1}{g} ( Q_5 \mu_5 V + Q_1 \mu_1 )  \label{tension}
\end{equation}
with $\mu_5 = (2\pi)^{-5} \alpha'^{-3}$ and $\mu_1 = (2\pi)^{-1}
\alpha'^{-1}$.

Now imagine taking $Q_1 < 0$, but keeping the {\it same} unbroken
supersymmetry.  This is not the same as replacing the D1-branes with
anti-D1-branes, which would leave unbroken $({\bf 1},{\bf 2})$
supersymmetries instead.  Rather, the solution~(\ref{15sol}) is simply
continued to $Q_1 < 0$, so that $r_1^2 < 0$. This radically changes
the geometry: the radius $r = |r_1|$ is now a naked
singularity~\cite{rep1,rep2,rep3}, and the region $r < |r_1|$ is
unphysical.  Also, the tension~(\ref{tension}) can vanish and
apparently even become negative.

In spite of these odd properties, the case $Q_1 < 0$ can be realized
physically.  To do this, replace the $T^4$ with a K3, with $Q_5$
D5-branes wrapped on the K3.  Then as shown in ref.~\cite{BSV}, the coupling
of the D5-brane to the curvature induces a D1-brane charge $Q_1 = - Q_5$.
For $g Q_5$ sufficiently large the solution~(\ref{15sol}) with $Q_1 = - Q_5$
(and with $dx^i dx^i$ replaced by the metric of a K3 of volume $V$) would be
expected to be a good description of the geometry.

The low energy theory on the branes is pure supersymmetric Yang--Mills
with eight supersymmetries~\cite{BSV}, in $1+1$ dimensions.  One can
understand this from the general result that the number of
hypermultiplets minus vector multiplets is $n_H - n_V = Q_5 Q_1$, from
1-5 strings.  Continuing to $Q_1 = - Q_5$ gives $n_H - n_V= -Q_5^2$,
corresponding to the $U(Q_5)$ adjoint without
hypermultiplets.\footnote{For larger $Q_1$, we can also study the case
of $SU(N)$ with $N_f$ fundamental flavours.  It is amusing to note
that the case $N_f=2N$ corresponds to $Q_1=0$, which means that
$Z_1=1$ and the supergravity solution simplifies greatly.  This is
pertinent for the four-dimensional case which is superconformal.}

By dualities one can find many other brane configurations with
singularities of the same sort, and with the same low energy gauge
theory in $p + 1$ dimensions.  Our interest in this solution arose
from the search for supergravity duals to these gauge theories, and we
will return to this point in section~4.  By $T$-dualities on the
noncompact directions one obtains solutions with D$p$ and D$(p+4)$
charge, for $p$ = 0, 1, 2, and 3.  By $T$-dualities on the whole K3
one can replace the D$(p+4)$-branes wrapped on the K3 with
D$(p+2)$-branes wrapped on a nontrivial $S^2$ with self-intersection
number $-2$.  This latter realization can also be obtained as follows.
Consider $N$ D3-branes at a $Z_2$ orbifold
singularity~\cite{z2}.  The low energy gauge theory is $U(N)
\times U(N)$ with hypermultiplets in the $({\bf N},\ol{\bf N})$.  By
going along the Coulomb branch in the direction diag$(I,-I)$ one gives
masses to all the $SU(N) \times SU(N)$ hypermultiplets.  This
corresponds to separating the branes along the singularity into two
clumps of $N$ half-branes, which are
secretly~\cite{joetensor,cvjrob,dougwrap} D5-branes wrapped on the
collapsed $S^2$ (the half D3-brane charge comes from the $\theta =
\pi$ $B$-field at the orbifold point).\footnote{This realization has
also been considered recently by E. Gimon and in refs.~\cite{igoretal}.
The latter consider $M$ wrapped D5-branes plus $N$ D3-branes, producing
gauge group $SU(N) \times SU(N+M)$ with bifundamental hypermultiplets.
The focus of these papers is $M \ll N$, whereas ours is the opposite limit
$N=0$.} A
$T$-duality on this brane-wrapped ALE space results in another dual
realization, this time  involving a pair of NS5-branes with $N$
D$(p+1)$-branes stretched between  them~\cite{hanany}.

Finally, by an $S$-duality the $p=0$ case can be related to the heterotic
string on $T^4$, with $N$ BPS winding strings having $N_L = 0$.  These are the
strings which become massless non-Abelian gauge bosons at special points in
moduli space, a fact that will play an important role in the next section.
Similarly,
the $p=2$ case $S$-dualizes~\cite{JKKM} to a combination of the
Kaluza--Klein  monopole and the
H-monopole~\cite{kkmono,hmono} which is equivalent~\cite{bound}
to the $a=1$ magnetic black hole in four dimensions.
In this heterotic form, these solutions have
previously been considered in refs.~\cite{rep1,rep2,rep3}.

The nature of the singularity was studied in ref.~\cite{rep2}.  It was shown
that massive particles coupled to the Einstein metric feel an infinite
repulsive potential at the singularity, hence the name {\it repulson}.
If instead one takes as probes the same kind of D5-brane as forms the
geometry, then by the usual supersymmetry argument the potential should
vanish.  Hence there should be no obstruction to building this geometry from
a collection of such D5-branes.
However, we will find in the next section that this argument fails for a
reason specific to the repulson geometry, so that in fact the geometry is
smoothed out in a certain way, and the singularity removed.

%--------------------------------------------------------------------+
\sect{The Enhan\c{c}on Geometry}

For $Q_1 < 0$ the gauge theory has no
Higgs branch (which would correspond to one or more bound states), and so for
$p
\leq 1$  infrared fluctuations might prevent the existence of a stable
object.  For this reason we will focus on the D2-D6 example in this
section,
\begin{eqnarray}
ds^2 \* Z_2^{-1/2} Z_6^{-1/2} \eta_{\mu\nu} dx^\mu dx^\nu +
Z_2^{1/2} Z_6^{1/2} dx^i dx^i + V^{1/2} Z_2^{1/2} Z_6^{-1/2} ds^2_{\rm K3} \
,\nonumber
\\
e^{2\Phi } \* g^2 {Z_2}^{1/2}{ Z_6}^{-3/2}\ , \nonumber\\
C_{\it 3} \* ({Z_2} g)^{-1} dx^0 \wedge dx^4 \wedge dx^5\ , \nonumber\\
C_{\it 7} \* ({Z_6} g)^{-1} dx^0 \wedge dx^4 \wedge dx^5 \wedge
dx^6 \wedge dx^7 \wedge dx^8 \wedge dx^9
\ . \label{26sol}
\end{eqnarray}
Now $\mu,\nu$ run over the 045-directions tangent to all the branes, $i$
runs over the 123-directions transverse to all branes, $ds^2_{\rm K3}$
is the metric of a K3 surface of unit volume, and
\begin{eqnarray}
Z_2 \* 1+\frac{r_2}{r}\ ,\quad  r_2 =
-\frac{ (2\pi)^4 g N \alpha'^{5/2} }{ 2V } \ , \nonumber\\
Z_6 \* 1+\frac{r_6}{r}\ ,\quad  r_6 = \frac{gN\alpha'^{1/2}}{2} \ ,
\end{eqnarray}
We have inserted $Q_2 = - Q_6=-N$.
The BPS bound for general charges is
\begin{equation}
\tau = \frac{1}{g} ( q_6 \mu_6 V + q_2 \mu_2 )  \label{tension2}
\end{equation}
with $\mu_6 = (2\pi)^{-6} \alpha'^{-7/2}$ and $\mu_2 = (2\pi)^{-2}
\alpha'^{-3/2}$.

A D6-brane probe (wrapped on the K3) has a well-defined
moduli space.  To understand better the physics of the repulson geometry,
consider the effective action of such a probe,
\begin{equation}
S = - \int_{M}d^3\xi\, e^{-\Phi(r)} (\mu_6 V(r) - \mu_2)
(-\det{g_{ab}})^{1/2} + \mu_6 \int_{M \times\rm K3}\! C_{\it 7} -  \mu_2
\int_{M} C_{\it 3}\ . \label{probe}
\end{equation}
Here $M$ is the projection of the world-volume onto the six noncompact
dimensions and $g_{ab}$ is the induced metric.  We have written this down
on physical grounds.  The first term is the Dirac action with the
position-dependence of the tension~(\ref{tension2}) taken into account;
in particular, $V(r)=V Z_2(r)/Z_6(r)$.  The second and third terms are
the couplings of the probe charges $(q_6,q_2) = (1,-1)$ to the background.
Note that to derive this action from the full D6-brane action requires two
curvature-squared terms.  One appears in the WZ action and accounts for
the induced D2 charge~\cite{BSV,aroof}.  The other appears in the Dirac
action and produces the $-\mu_2$ term in the tension~\cite{sunilone,BBG}.

Expanding the action~(\ref{probe}) in powers of the transverse velocity gives
the Lagrangian density
\begin{eqnarray}
{\cal L} \* - \frac{\mu_6 V Z_2 - \mu_2 Z_6}{Z_6 Z_2 g}  + \frac{\mu_6
V}{g}(Z_6^{-1} - 1) - \frac{\mu_2}{g} (Z_2^{-1} - 1)  \nonumber\\
&&\qquad\qquad\qquad\qquad{}+\frac{1}{2g}
(\mu_6 V Z_2 -
\mu_2 Z_6) v^2 + O(v^4)\ . \label{lag}
\end{eqnarray}
 The position-dependent potential terms cancel as expected for a
supersymmetric system, leaving the constant potential $(\mu_6 V - \mu_2)/g$
and a nontrivial metric on moduli space as expected with eight
supersymmetries.  The metric is proportional to
\begin{equation}
\mu_6 V Z_2 - \mu_2 Z_6 = (2\pi)^{-2} \alpha'^{-3/2} \Biggl(
\frac{V}{V_*} - 1 - \frac{g N \alpha'^{1/2}}{r}\Biggr)\ .\label{kinetic}
\end{equation}
We assume that $V > V_* \equiv (2\pi)^4 \alpha'^2$, so that the metric at
infinity (and the membrane tension) are positive.  However, as $r$
decreases the  metric eventually becomes negative, and this occurs at a
radius
\begin{equation}
r = \frac{2V}{{V} - V_* } |r_2| \equiv r_{\rm e}
\end{equation}
which is strictly greater than the radius $r_{\rm r} = |r_2|$ of the
repulson singularity.

To understand what is happening, note that this kinetic term comes
entirely from the Dirac term in the action~(\ref{probe}), and that what is
vanishing is the factor $\mu_6 V(r) - \mu_2$ in the probe tension.
This occurs when $V(r)=\mu_2/\mu_6 = V_*$, and so before the singularity
(where $V(r)$ goes to 0).  The negative tension at $r_{\rm e} > r > r_{\rm
r}$ is clearly unphysical.  To see how we should interpret it, recall that
the probe is dual to a heterotic winding string, and the vanishing of the
tension corresponds to the vanishing of the winding string mass at a point
of enhanced gauge symmetry.  It is well-known that the latter can be
interpreted as the ordinary Higgs mechanism.  In the Higgs mechanism the
mass is related to the expectation value by
\begin{equation}
m = \lambda |\phi|\ ;
\end{equation}
note the absolute value.

Therefore we should take (minus) the absolute
value in the first term of the action~(\ref{probe}) and (\ref{lag}).
Now the metric is positive but another problem appears: the potential no
longer cancels, but rises as $r$ decreases below $r_{\rm e}$.  This
means that we cannot move the probe to $r < r_{\rm e}$ in a supersymmetric
way, and so contradicts the assumption that we can build the repulson
geometry by bringing together a succession of wrapped D6-branes.  Thus we
are led to a very different picture: the $N$ D6-branes all live on the
sphere at $r = r_{\rm e}$.  Even if we try to build the geometry by
starting with coincident D6-branes at $g=0$, where $r_2 = r_{\rm r} =
r_{\rm e} = 0$, and then increasing the coupling, we would expect the
system to expand as $g$ is increased.  We will see an interesting parallel
to this behavior in the gauge theory discussion of the next section.

With this picture, the geometry~(\ref{26sol}) is correct only down to $r =
r_{\rm e}$.  Since the sources are all at this radius, the geometry should
be flat in the interior $0 < r < r_{\rm e}$:\footnote{
We assume that $V = V_*$ in the interior by continuity of the metric.  C.
Vafa suggests that there may be an `overshoot,' by analogy with
ref.~\cite{gova}.}
\begin{eqnarray}
ds^2 \* [Z_2(r_{\rm e}) Z_6(r_{\rm e})]^{-1/2} \eta_{\mu\nu} dx^\mu
dx^\nu + [Z_2(r_{\rm e}) Z_6(r_{\rm e})]^{1/2} dx^i dx^i
+ V_*^{1/2} ds^2_{\rm K3} \ ,\nonumber
\\
e^{2\Phi } \* g^2 {Z_2(r_{\rm e})}/{ Z_6(r_{\rm e})}\ , \nonumber\\
C_{\it 3} \*
C_{\it 7} = 0
\ . \label{26flat}
\end{eqnarray}
Note that the non-zero potential that appeared in the
proof-by-contradiction is not a real feature, because the
geometry~(\ref{26sol}) is no longer relevant for $r < r_{\rm e}$.  Indeed,
it is difficult to see how such a potential could be consistent with
supersymmetry.

Now, however, we seem to have another contradiction.  There seems
to be no obstacle to the probe moving into the flat
region~(\ref{26flat}), contradicting the conclusion that the D6-branes are
fixed at $r_{\rm e}$.  To see the obstacle we must look more deeply.
Note that in the interior region the K3 volume takes the constant value
$V_*$, meaning that the probe is a tensionless membrane.  A tensionless
membrane sounds even more exotic than a tensionless string, but in fact
(as in other examples) it is actually prosaic: it is best interpreted as a
composite in an effective field theory.  Note that the ratios 
$\mu_0/\mu_4{=}\mu_2/\mu_6{=}V_*$ 
are equal.  This means that a wrapped D4-brane is a
massless particle whenever the wrapped D6-brane is
tensionless.  In fact, it is a non-Abelian gauge boson, which together with
an R--R vector and a wrapped anti-D4-brane form an enhanced $SU(2)$ gauge
symmetry.  That is, in the interior geometry there is an unbroken $SU(2)$
gauge symmetry in six dimensions.  For this reason we refer to this as the
enhan\c{c}on geometry, and the radius $r_{\rm
e}$ as the enhan\c{c}on radius.

Now, a two-dimensional object in six dimensions would be obtained by
lifting a point object in four, and so a magnetic monopole naturally
suggests itself.  Indeed, the wrapped D4-brane is a source of a 2-form
R--R field strength in six dimensions, and the wrapped D6-brane is the
source of the dual 4-form field strength.  The mass of a monopole is
proportional to the mass of the corresponding $W$ boson, so they vanish
together when $V(r) = V_*$.  Since the size of the monopole is inverse to
the mass of the $W$ boson, there is no sense in which a probe can be
localized within the enhan\c{c}on radius.  For this same reason the probe
begins to expand as it approaches $r_{\rm e}$, so it appears that
it will essentially melt smoothly into the shell of
$N$ monopoles at $r_{\rm e}$.  This has the effect that junction between
the exterior geometry~(\ref{26sol}) and the interior geometry~(\ref{26flat})
is smoothed.

We can estimate this smoothing effect as follows.  The mass of a
wrapped D-brane is $m(r) = e^{-\Phi} \mu_4 (V-V_*)$.
The probe will cease to be effectively pointlike when
\begin{equation}
m(r) (r - r_{\rm e})g_{rr}^{1/2} \sim 1\ ,
\end{equation}
leading to
\begin{equation}
(r - r_{\rm e}) \sim r_{\rm e} N^{-1/2}\ .
\end{equation}
Thus we have a consistent picture in
which the repulson is replaced by a smooth geometry.\footnote{R. Myers and
A. Strominger suggest that this may apply to more general
Reissner--Nordstrom-like singularities.}

The same principle holds for other values of $p$.  The
enhan\c{c}on locus is $S^{4-p} \times
R^{p+1}$, whose interior is $(5+1)$-dimensional.  For even $p$ the theory
in the interior has an $SU(2)$ gauge symmetry, while for odd $p$ there is
an $A_1$ (2,0) theory.  This is consistent with the fact that a K3 with
volume $V_*$ is $T$-dual to a K3 at an $A_1$ singularity.  The details of
the smoothing depend on $p$, and for $p \leq 1$ it is likely that the IR
fluctuations must be considered.\footnote{We thank S. Sethi for discussions
on this point.}

Note that our result for the Lagrangian density (\ref{lag}) depends
only on three moduli space coordinates, $(x^3,x^4,x^5)$, or
$(r,\theta,\phi)$ in polar coordinates.  For a (2+1)-dimensional theory with
eight supercharges, the moduli space metric must be
hyperK\"ahler~\cite{hyper}. A minimum requirement for this is of course that
it has four coordinates, and so we must find an extra modulus. On the probe,
there is an extra $U(1)$ gauge potential $A_a$, corresponding to the overall
centre of mass degree of freedom. We may exchange this for a scalar $s$ by
Hodge duality in the (2+1)-dimensional world-volume. This is of course a
feature specific to the $p = 2$ case.

To get the coupling for this extra modulus correct, we should augment
the probe computation of the previous section to include $A_a$. The
Dirac action is modified by an extra term in the determinant:
\begin{equation}
-{\rm det}g_{ab}\to-{\rm det}(g_{ab}+2\pi\alpha^\prime F_{ab})\ ,
\end{equation}
where $F_{ab}$ is the field strength of $A_a$. Furthermore, in the
presence of $F_{ab}$, there is a coupling
\begin{equation}
-2\pi\alpha^\prime\mu_2 \int_M C_1 \wedge F\ ,
\end{equation}
where $C_1=C_\phi d\phi$ is the magnetic potential produced by the
D6-brane charge: $C_\phi={-}(r_6/g)\cos\theta$.
Adapting the procedures of refs.~\cite{townsend,schmidhuber}, we can
introduce an auxiliary vector field $v_a$, replacing
$2\pi\alpha^\prime F_{ab}$ by $e^{2\phi}(\mu_6V(r)-\mu_2)^{-2}v_av_b$
in the Dirac action, and adding
 the term $2\pi\alpha^\prime\int_M F\wedge v$
overall. Treating $v_a$ as a Lagrange multiplier,
the path integral over $v_a$ will give the action involving $F$ as before.
Alternatively, we may treat $F_{ab}$ as a
Lagrange multiplier, and integrating it out enforces
\begin{equation}
\epsilon^{abc}\partial_b(\mu_2 {\hat C}_c+v_c)=0\ .
\end{equation}
Here, ${\hat C}_c$ are the components of the pullback of $C_1$
to the probe's world-volume.
The solution to the constraint above is
 \begin{equation}\mu_2{\hat C}_a+v_a=\partial_a s\ ,
\end{equation} where the  scalar $s$ is
our fourth modulus. We may now replace $v_a$ by $\partial_a
s-\mu_2{\hat C}_a$ in the action, and the static gauge computation
gives for the kinetic term:
\begin{equation}
{\cal L}=
F(r)
 \left({\dot r}^2
 +r^2{\dot\Omega}^2 \right)
+F(r)^{-1}\left({\dot s}/2-\mu_2C_\phi{\dot\phi}/2\right)^2\ ,
\label{fourth}
\end{equation}
where
\begin{equation}
F(r)={Z_6\over 2g}\left(\mu_6V(r)-\mu_2\right)\ ,
\end{equation}
and ${\dot\Omega}^2={\dot\theta}^2+\sin^2\!\theta\,{\dot\phi}^2.$

%--------------------------------------------------------------------+
\sect{Gauge Theory}

\subsection{The Search for a Duality}

One of the goals of this work is to obtain a useful dual description
of the physics of strongly coupled $SU(N)$ gauge theory (with eight
supercharges and no hypermultiplets) at large-$N$.  This is a
necessarily complex undertaking, as there are at least four different
theories which play important roles here, and so in the spirit of
ref.~\cite{IMSY}, we should carefully determine where each theory has a
weakly coupled description, as we change the energy scale.

To get to the limit where we obtain the decoupled gauge theory we hold
fixed the induced $p$-dimensional gauge coupling
\begin{equation}
g_{{\rm YM},p}^2 = (2\pi R)^{-4} g_{p+4}^2 = (2\pi)^{p-2} g
\alpha'^{(p+1)/2}R^{-4}
\end{equation}
and, as usual~\cite{juan}, hold fixed $U=r/\alpha'$.  Let us also
define the $(p+1)$-dimensional 't Hooft coupling
\begin{equation}
\lambda_p \equiv g_{{\rm YM},p}^2 N \,
\end{equation}
where $N$ is the number of D$(p+4)$-branes wrapped on the K3. We
write the K3 volume as $V \equiv (2\pi R)^4$;
the background has a good limit if we hold this fixed as well.  Then the
string metric becomes, in the decoupling limit,
\begin{equation}
\begin{array}{rl}\vspace{5pt}
{\displaystyle{ {{ds^2}\over{\alpha'}}}} = &\  \
\left[h_p(U)\left(1-h_p(U)\right)\right]^{-1/2}
R^{-2}\eta_{\mu\nu}dx^\mu dx^\nu 
   \\\vspace{5pt}
 &\! +
\left[\left(1-h_p(U)\right)/h_p(U)\right]^{1/2}  (2\pi)^2 ds^2_{\rm K3}
   \\
 &\! + \left[h_p(U)\left(1-h_p(U)\right)\right]^{1/2}
  R^2\left(dU^2+U^2d\Omega_{4-p}^2\right)
\,.
\label{scaledmetric}
\end{array}\end{equation}
We have abbreviated $c_p$ as
$(2\sqrt{\pi})^{5-p}\Gamma\left((7-p)/2\right)$, and
\begin{equation}
h_p(U)={{c_{p+4}}\over{(2\pi)^{p-2}}} {{\lambda_p}\over{U^{3-p}}}
\end{equation}
The dilaton becomes
\begin{equation}
\label{dilp}
e^\Phi = {{\lambda_pR^{3-p}}\over{(2\pi)^{p-2}N}}
\left(h_p(U)\right)^{-(p+1)/4}\left(1-h_p(U)\right)^{(3-p)/4}\ .
\end{equation}
Note that the `1' has scaled out of $Z_{p+4}$ but not $Z_p$.

The case of $p = 3$ needs to be discussed separately.
The spacetime solution is
\begin{eqnarray}\label{peq3spacetime}
ds^2 \* Z_3^{-1/2} Z_7^{-1/2} \eta_{\mu\nu} dx^\mu dx^\nu +
Z_3^{1/2} Z_7^{1/2} (\alpha^\prime)^2 dud{\bar{u}}
+ (2\pi R)^2 Z_3^{1/2} Z_7^{-1/2} ds^2_{\rm K3} \
,\nonumber\\
e^{\Phi } \* g {Z_7}^{-1}\ , \nonumber\\
Z_3 \* \frac{g N}{2\pi}\frac{(\alpha^\prime)^2}{R^4} \ln{(U/\rho_3)}
 \ , \nonumber\\
Z_7 \* \frac{g N}{2\pi} \ln{(\rho_7/U)}\ ,
\end{eqnarray}
where $U=|u|$.
This is sensible only for $\rho_3 < U < \rho_7$.
At $U = \rho_3$, the inner radius, $Z_3$ vanishes:
this is the repulson, which is again unphysical, lying inside the
enhan\c con.  At $U = \rho_7$, the outer radius, $Z_7$ vanishes and
the dilaton diverges. Near this radius there is a story similar to
that at the inner radius: nonperturbative corrections remove the
singularity.  For $N \leq 24$, this is understood in terms of
(choke) F-theory~\cite{F}.  For $N > 24$, as here, the details are
not so well understood, but should not be relevant to the physics in
the enhan\c con region.

Returning to $p<3$, can work out where the supergravity description
(\ref{scaledmetric},\ref{dilp}) is good by demanding~\cite{IMSY} that
the curvature in string units and the dilaton both be small. For the
curvature, we find
\begin{equation}
\alpha' {\cal{R}} = -g_p(U) U^{-(p+1)/2} (\lambda_p R^4)^{-1/2}
\times \left(1-h_p(U)\right)^{-5/2}\ ,
\end{equation}
where the functions $g_p(U)$ are ${\cal{O}}(1)$ for all $U\ge{U_{\rm e}}$.

The first thing we notice about the curvature is its value at the
enhan\c con radius:
\begin{equation}
\left. \alpha'{\cal{R}} \right|_{\rm e} \sim
(\lambda_p R^{3-p})^{-2/(3-p)} \,.
\end{equation}
The control parameter
\begin{equation}
\mu \equiv \lambda_p R^{3-p}
\end{equation}
will determine the nature of the phase diagram.  The physical
interpretation of $\mu$ is the value of the dimensionless 't Hooft
coupling of the $(p+1)$-dimensional gauge theory at the energy scale
$1/R$, which  is an effective UV
cutoff.  Below this scale, since the physics is superrenormalizable,
the effective coupling grows, becoming strong at $E < \lambda^{-1}$.
At this point the gauge theory ceases to be a useful description, we
have the right to look for a supergravity (or other) dual.  If
$\lambda_p$ is small, we expect to find a region of the phase diagram
where gauge theory is a weakly coupled description.  Otherwise, we
will have only supergravity phases.  The interesting case is therefore
$\mu \ll 1$, so that we have at least one region where the gauge
theory is weakly coupled.  We will take
\begin{equation}
 \label{weak}
\lambda_p R^{3-p}\ll 1
\end{equation}
for the remainder of this subsection.

In satisfying this condition, we find that the supergravity geometry
is strongly curved at the enhan\c con radius.  Since the supergravity
fields do not evolve inside the enhan\c con, this is the maximum
curvature.  At the enhan\c con radius, we can also inspect the dilaton;
it is
\begin{equation}
\left. e^\Phi \right|_{\rm e} \sim {{\lambda_p R^{3-p}}\over{N}} \ll 1 \,.
\end{equation}
{}From the equation (\ref{dilp}), we find that the dilaton increases
monotonically with $U$.  It becomes of order one at
\begin{equation}
U_2^{3-p} \sim \lambda_p  N^{4/(p+1)}
(\lambda_p R^{3-p})^{-4/(p+1)}
\end{equation}
At radii $U>U_2$, we will need to use the $S$-dual supergravity
description.  Since at these radii the effect of $Z_p$ is very small,
by comparison to the effect of $Z_{p+4}$, we are in fact matching on
to the picture obtained by~\cite{IMSY}  for the D$(p+4)$-branes alone;
the physics involving the K3 is essentially irrelevant.

We now need to find the degrees of freedom best suited to describing
the physics for $U_{\rm e}<U<U_2$.  As $U$ decreases from $U_2$, the
curvature of the $d=10$ supergravity geometry will become stronger; it
becomes of order unity at
\begin{equation}
U_1^{p+1} \sim 1/ \lambda_p R^4
\end{equation}
At this place, the $(p+5)$-dimensional gauge theory will take over.
We can see this by starting with its dimensionless 't Hooft coupling
as a function of gauge theory energy $E$
\begin{equation}
\lambda_{p+4} E^{p+1}  \,.
\end{equation}
Now, in order to relate $E$ to $U$ we need an IR/UV relation for the
strings stretched between the  probe brane and the source branes.  
Since the source branes are distributed on a
S$^{4-p}$ shell at the enhan\c con radius, the relation is $E=U-\eta
U_{\rm e}$, where $-1\le\eta\le 1$ encodes which brane on the shell the
stretched string attaches to.  In addition, we have the condition on
the control parameter $\mu\ll 1$, so that $U_1, 1/R \gg U_{\rm e}$ and 
thus the
size of the shell is not important.  So the IR/UV relation is to good
accuracy $E=U_{\rm e}$.  Therefore we see that the dimensionless 't Hooft
coupling of the $(p+5)$-dimensional gauge theory is order one at $U_1$.
It decreases for smaller $U$.

We also have the induced lower-dimensional gauge theory with
dimensionless coupling
\begin{equation}
\lambda_p E^{p-3}\ .
\end{equation}
The crossover between the $(p+5)$- and $(p+1)$-dimensional gauge theories 
is clearly at $E_*=1/R$.  At lower energies {\it i.e.}\ smaller $U$, the
$(p+1)$-dimensional gauge theory takes over.  However, this time its
dimensionless 't Hooft coupling increases as $U$ decreases,
{\it i.e.} exhibits the opposite behavior to the $(p+5)$-dimensional gauge
theory.  This is consistent with the physics of the ${\cal{N}}=4$
supersymmetric systems studied in~\cite{IMSY}.  The 
dimensionless 't Hooft coupling of the $(p+1)$-dimensional theory
becomes strong at
\begin{equation}
E_0\sim\lambda_p^{1/(3-p)}
\end{equation}
Let us now recall the enhan\c con radius, defined by $h(U)=1/2$, or
$U^{3-p} \sim 2\lambda_p$.  From our previous considerations of a
probe in the background of $N$ branes, we found that the branes were
(evenly) distributed on a S$^{4-p}$ shell of this radius.  The energy
of a string stretched between any two branes in this shell must
therefore fall between
\begin{equation}
\label{newrange}
(\lambda_p)^{1/(3-p)}/N^{1/(4-p)} < E < (\lambda_p)^{1/(3-p)}
\end{equation}
The energy at which the $p$-dimensional gauge theory becomes strong is
then precisely the energy at which the physics of the system is
described by some theory whose dynamics includes only the BPS strings
stretched between the source branes (and of probe branes so close to
the enhan\c con radius that they too can be thought of as source
branes).

At the very lowest energies we can simply use the moduli space
description of the physics.

The gap that remains in building our phase diagram is an understanding
of the physics in the energy range (\ref{newrange}).  The curvature is
strong there and thus we fail to find a supergravity dual for the
strongly coupled $(p+1)$-dimensional gauge theory.  A further
indication of that failure is the explicit appearance of $R$ in the
metric~(\ref{scaledmetric}): in the gauge theory, $R$ enters the physics
only through the parameter $g^2_{{\rm YM},p}$, and therefore should not
explicitly appear in a dual description.

So we have a mystery here. A natural suggestion for $p=2$ (with a
suitable generalization involving the $A_1$ theory for $p=3$) is that
the dual is the (5+1)-dimensional $SU(2)$ gauge theory in the
$N$-monopole sector, where they are just becoming massless.  Note that
this $SU(2)$ gauge theory is part of the bulk physics, so what we are
conjecturing is that it is the only relevant part and that the
supergravity can be omitted.  A weak test is that it give the correct
moduli space, and it does.  Note also that the enhan\c con geometry
has a natural interpretation in the gauge theory: in the $N$-monopole
sector the Higgs field has a zero of order~$N$.  The function $r^N$ is
essentially zero (the flat interior) until rising sharply.  Thus a
classical monopole solution of large charge might be expected to have
its charge distributed in a thin shell.  To go further we need a new
expansion to describe this system, which becomes weak for $N$ large.
It appears that the spacing of the monopoles, of order $N^{-1/2}$
times the enhan\c con radius, plays the role of a
``non-commutativity'' parameter, because the sphere has been
effectively broken up into $N$ domains.  In any case, the mysterious
dual description should include the dynamics of these stretched
strings at large-$N$.

Again, the case $p=3$ needs special treatment.  To orient ourselves,
we may review the case of D7-branes by themselves.  Using the
equations (\ref{peq3spacetime}), we find that the scalar curvature in
string units is
\begin{equation}
\alpha^\prime {\cal{R}} \sim \left[\sqrt{N}\alpha^\prime U^2
\left(\log(\rho_7/U)\right)^{5/2} \right]^{-1} \,,
\end{equation}
and this becomes order unity when $U=U_1\sim (\alpha^\prime)^{-1/2}
(gN)^{-1/4}$.  Substituting this into the dimensionless 't Hooft
coupling on the D7-branes, we find that the gauge coupling is order
unity at the same place, and so the $(7+1)$-dimensional gauge theory
and ten-dimensional supergravity parts of the phase diagram fit
together as required. The dilaton becomes order unity when
$U=\rho_7e^{-2\pi/N}\sim \rho_7$, and because $N>24$, the nature of
this theory beyond the supergravity approximation is unclear.

Let us now add the D3-branes.  The effect on the curvature is
essentially to multiply the above result by a factor $Z_3^{-1/2}$.
Out at large values of $U$ such as $U_1$ and $\rho_7$, the effect of
the $Z_3$-factor must be unimportant, by analogy with the lower-$p$
cases.  
Now, recall that at the scale
$1/R$, the $(7+1)$-dimensional gauge theory crosses over to the
$(3+1)$-dimensional gauge theory, and the coupling must be weak
there in order for there to be gauge theory descriptions at
all. Therefore $\lambda_3\ll 1$, and as a consequence we find
$\rho_3\ll 1/R$, a condition necessary for
clean separation of some of the phases as in the lower-$p$ cases.
The remaining conditions needed for clean separation of the other
phases involve $\rho_7$ and remain somewhat puzzling.  
We assume, by analogy with the $p<3$ cases, that they are met.

\begin{figure}
\epsfysize=3.5in
\hskip0.2\textwidth\epsfbox{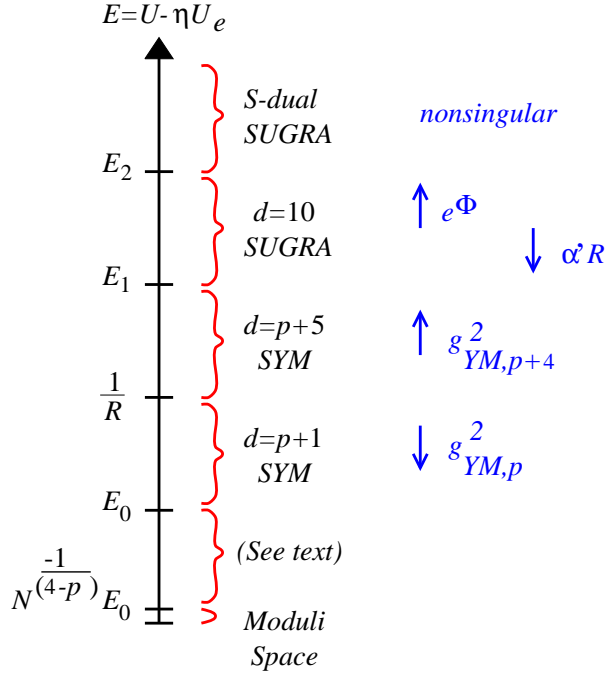}
\caption{\small
The phase diagram for large-$N$ and $\lambda_p R^{3-p}\ll{1}$.  Note
that when $\lambda_p R^{3-p} \gg 1$, the supergravity plus $SU(2)$
gauge description in section~3 should be valid at all radii below
$U_2$: the two SYM phases and the mystery phase disappear.  }
\label{fig:phases}
\end{figure}

We summarize our findings in the phase diagram in figure~\ref{fig:phases}.

To get a little more information on the mystery theory we may consider
going to finite temperature, or adding energy to the branes.  The
first change to the supergravity solution is that a nonextremality
function $k(U)$ appears multiplicatively in $g_{tt}, g_{UU}^{-1}$:
\begin{equation}
k(U) = 1-\left({{U_0}\over{U}}\right)^{3-p}\,.
\end{equation}
In addition, the harmonic function of the D$p$-branes $Z_p$ gets
altered by nonextremality as if we had not taken the decoupling limit,
while $Z_{p+4}$ is unaltered\footnote{The reason that the branes are
affected asymmetrically by nonextremality is that we held $R$ fixed in
the decoupling limit; it does not scale with the string length.}
Specifically,
\begin{equation}
Z_p \rightarrow 1- {{\xi c_{p+4}\lambda_p}\over{(2\pi)^{p-2}U^{3-p}}} \,,
\end{equation}
where
\begin{equation}
\xi = \left[1+ 
\left({{(2\pi)^{p-2}U_0^{3-p}}\over{2c_{p+4}\lambda_p }}\right)^2 
\right]^{1/2}
- {{(2\pi)^{p-2}U_0^{3-p}}\over{2c_{p+4}\lambda_p }} \,.
\end{equation}
Note that $1\ge\xi>0$, and therefore the enhan\c con 
at $U=[(1+\xi)/2]U_{\rm e}$ is pulled
inwards to smaller $U$ as the ratio $U_0^{3-p}/\lambda_p$ is
increased.

In order for the supergravity horizon at $U=U_0$ to lie outside the
enhan\c con locus, we need a sufficiently large energy density
$\varepsilon$ on the branes.  Using the relation \cite{IMSY}
$U_0^{3-p} \sim g_{p+4}^4\varepsilon$, this implies that
\begin{equation}
\varepsilon R^{p+5} > {{N^2}\over{\lambda_p R^{3-p}}} \gg 1 \,,
\end{equation}
where we have used the condition (\ref{weak}).  This cannot be
satisfied in the $(p{+}1)$-dimensional gauge theory, and this is
further evidence that the mysterious theory is not gravitational.

%--------------------------------------------------------------------+

\subsection{The Metric on Moduli Space}

Although we have failed to find a weakly coupled dual for the gauge
theory, a study of the Coulomb branch of the gauge theory reveals a
striking parallel to our probe computations.  Focusing on the case
$p = 2$, suppose that the condition~(\ref{weak}) holds, so the low
energy physics can be described by a $(2+1)$-dimensional gauge
theory.  The metric that the probe sees, after applying the scaling of
the previous section to eqn.~(\ref{fourth}), gives the metric on
moduli space for the $SU(N)$ (2+1)-dimensional gauge theory in the
decoupling limit:
\begin{equation}
{\cal L}= f(U) \left({\dot U}^2 +U^2{\dot\Omega}^2\right) +f(U)^{-1}
 \left({\dot\sigma} -{N\over{8\pi^2}}A_\phi{\dot\phi}\right)^2\
 ,\label{probenear}
\end{equation}
where
\begin{equation}
f(U)={1\over 8\pi^2 g^2_{\rm YM}}
\left(1-{ \lambda \over U}\right)\ ,
\end{equation}
the $U(1)$ monopole potential is $A_\phi=\pm1-\cos\theta,$ and
$\sigma=s{\alpha^\prime}/2$, and the metric is meaningful only for
$U{>}U_{\rm e} = \lambda$.  Our metric (\ref{probenear}) is the
Euclidean Taub--NUT metric, with a negative mass.  It is a
hyperK\"ahler manifold, because $\nabla f=\nabla{\times}A$, where 
$A=(N/8\pi^2)A_\phi d\phi$.

It is striking that the moduli space metric
is the same as that which can be derived from
field theory, where this has the interpretation as the tree-level plus
one-loop result.  It is also interesting to note that while the
scaled supergravity solution failed to give a dual, the result for the
probe's moduli space is independent of $V$.  Both of these facts could
be understood if supersymmetry prevented $V$ from appearing in the
probe metric: one can use $V$ as the control parameter to move from
weak gauge theory to weak supergravity.  On the surface this does not seem
to be the case --- the probe moduli live in a vector multiplet, and so
does $V$ --- but a more careful analysis may be needed.
So there is a mystery
here, perhaps confirming our suggestions in the previous section that there
is a useful duality to be found.

We see that the enhan\c{c}on phenomenon is the same as the familiar
fact that the tree-level plus one-loop kinetic term goes negative
--- the Landau pole.  This metric
is of course singular, and is therefore incomplete.    As shown by Seiberg
and Witten, it receives no perturbative corrections but is fixed
nonperturbatively.  It is the large
$r$ expansion of the metric on the moduli space of $N$
monopoles. There are nonperturbative instanton corrections to this
metric which smooth it out into a generalization of the
Atiyah--Hitchin manifold~\cite{atiyah}. In the two-monopole case
studies in ref.~\cite{SWtwo}, the Atiyah--Hitchin manifold is the
unique smooth completion of the Taub--NUT metric consistent with the
condition of hyperK\"ahlerity. (See refs.~\cite{SWtwo,CH,hanany,khoze}
for examples of generalizations and further study.)

To discuss quantitatively the nonperturbative corrections to the metric on
moduli space it is simpler to look at the case $p=3$.
On the gauge theory side, the metric on moduli space is obtained from the
Seiberg--Witten curve~\cite{SW}, which for $SU(N)$ is~\cite{SWn}
\begin{equation}
y^2 = \prod_{i=1}^N (x - \phi_i)^2 - \Lambda^{2N}\ .
\end{equation}
The point of maximal unbroken gauge symmetry, which in the present
case can only be the Weyl subgroup, is
\begin{equation}
\phi_i = 0\ ,\quad \mbox{all $i$}\ .
\end{equation}
Earlier work on the
large-$N$ limit~\cite{DS} focused on a different point in moduli
space, but this highly symmetric point would seem to be the most natural
place to look for a supergravity dual.
The branch points $y=0$ are at
\begin{equation}
x = \Lambda e^{i\pi k/N}\ ,\quad 0 \leq k \leq 2N-1\ . \label{ring}
\end{equation}
This ring of zeros is reminiscent of the enhan\c con, and is in fact
the same.  To see this add a probe brane at $\phi$,
\begin{equation}
y^2 = x^{2N} (x - \phi)^2 - \Lambda^{2N+2}\ . \label{poly}
\end{equation}
For $|\phi| > \Lambda$, there are $2N$ zeros which closely approximate
a ring,
\begin{equation}
x \sim \Lambda e^{i\pi k/N} ( e^{i\pi k/N} - \phi/\Lambda)^{-1/N}
\ ,\quad 0 \leq k \leq 2N-1 \ ,  \label{outer}
\end{equation}
the new factor, in parentheses, being $1 + O(1/N)$.
The remaining two zeros are at $x \sim \phi$, or more precisely
\begin{equation}
x \sim \phi \pm (\Lambda/\phi)^{2N}\ ,
\end{equation}
the correction being exponentially small.  On the other hand, for
$|\phi| < \Lambda$, all $2N + 2$ branch points lie approximately in a
ring,\footnote{Inserting either form~(\ref{outer}) or (\ref{inner}) into the
polynomial (\ref{poly}) produces a solution to order $1$, which can be
further improved by an $O(N^{-2})$ correction to $x$.  The difference
between the two ranges of $\phi/\Lambda$ is that the terms in
parentheses are arranged so as not to circle the origin, so that the $1/N$
root comes back to its original value  as $k$ increases
from zero to $2N-1$ or $2N+1$.}
\begin{equation}
x \sim \Lambda e^{i\pi k/(N+1)} (1 - \phi e^{-i\pi k/(N+1)}/\Lambda)^{-1/N}
\ ,\quad 0 \leq k \leq 2N+1 \ . \label{inner}
\end{equation}
As deduced from the
string picture, the probe does not penetrate the interior of the
enhan\c con but rather melts into it.

One can extract the moduli space metric from the usual
formalism~\cite{SW,SWn}.  We leave a more detailed treatment for the future,
but in the probe region $\phi > \Lambda$ it matches the perturbative result
\begin{equation}
ds^2 \sim N \ln \Biggl(\frac{U}{\Lambda}\Biggr) du d\bar u\ .
\end{equation}
This agrees with the D3-brane probe result on the supergravity side
if we identify $\Lambda = \rho_{\rm e} = \sqrt{\rho_3 \rho_7}$.  

%--------------------------------------------------------------------+
\sect{A Tale of Two Duals}
As mentioned in the introduction, there are a number of $T$- and $S$-
dual pictures where the same physics arises. (The physics of the
heterotic $S$-dual for the $p = 2$ case is essentially contained in the
recent work of ref.~\cite{morten}.)

\begin{figure}
\hbox{\epsfxsize=2.2in\epsfbox{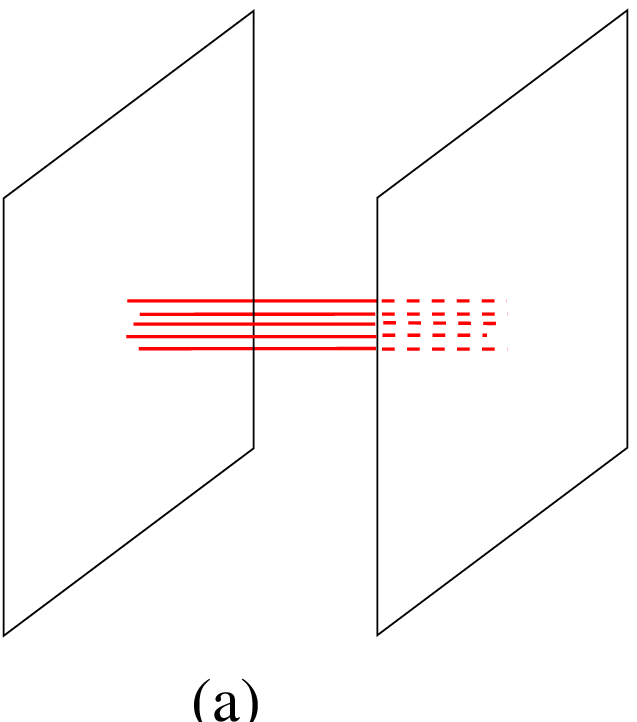}\hskip1in
\epsfxsize=2.2in\epsfbox{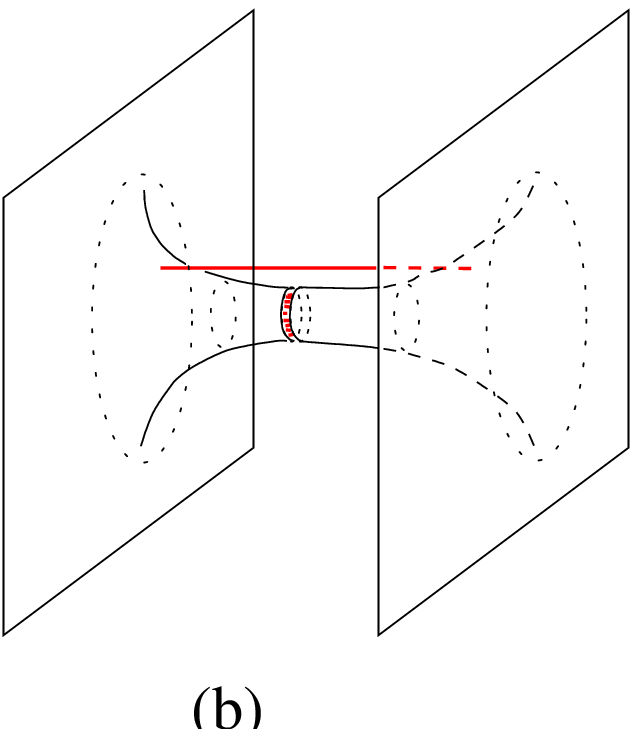}}
\caption{\small
$N$ D$(p+1)$-branes ending on NS5-branes: (a) The
classical picture (b) The corrected picture, showing the resulting
bending of the NS5-branes for large $gN$. The separated brane is
the probe which becomes massless at the enhan\c con locus, an
$S^{4-p}$ (a circle in the figure).}
\label{bendy}
\end{figure}

To construct some $T$-dual cases, consider~\cite{hanany} a pair of
NS5-branes which are pointlike in the $(x^6,x^7,x^8,x^9)$ directions,
with $N$ D$(p+1)$-branes stretched between them along the $x^6$
direction, where $p=0,1,2$, or 3. The latter are pointlike in the
$(x^{p+1},\ldots,x^5)$ directions, which are inside the NS5-branes,
and also in the $(x^7,x^8,x^9)$ directions. All of the branes share
the directions $(x^0,\ldots,x^p)$. This arrangement of branes, shown
in figure
\ref{bendy}(a) preserves eight supercharges. We will state the
general $p$ case in many of the following formulae. The reader may
wish to keep the $p = 2$ case in mind for orientation.

Denoting the separation of the NS5-branes in the $x^6$ direction
by $L$, there is a $(p + 1)$-dimensional $SU(N)$ gauge theory on the
infinite part of the world-volume of the D$(p + 1)$-branes, whose
coupling is
\begin{equation}
g^2_{\rm YM}=L^{-1}g^2_{p+1}=L^{-1}(2\pi)^{p-1}g\alpha^{\prime(p-2)/2}\ .
\label{coupling}
\end{equation}

A number of gauge theory facts from the previous sections are manifest
here. For example, the fact that (for $p = 2$) the Coulomb branch of
the $SU(N)$ gauge theory is dual to the moduli space of $N$
monopoles of a (5+1)-dimensional gauge theory follows from the fact
that the ends of the D3-branes are membrane
monopole sources (in $x^3,x^4,x^5$) in the NS5-branes' world-volume theory.
This is an $SU(2)$ gauge theory spontaneously broken to $U(1)$ by the
NS5-branes' separation. This will always be the relevant (5+1)-dimensional
theory when $(p + 1)$ is odd because we are in type~IIB string theory.  When
$(p + 1)$ is even, we are in type~IIA, and the (5+1)-dimensional theory is
the $A_1$ (0,2) theory.

Now place the $x^6$ direction on a circle of radius $2\pi\ell$. There
is a $T_6$-dual of this arrangement of branes.\footnote{See
refs.~\cite{cvjfrom,geometric,suniltwo} for discussions of dualities
of this sort.}  The NS5-branes become an $A_1$ ALE
space~\cite{teesix}.  To see this in supergravity language, we start
by smearing the NS5-branes along the $x^6$ space, writing the
supergravity solution for the core of the NS5-branes as~\cite{harvey}
($|{\bf y}|\gg\ell$):
\begin{eqnarray}
ds^2&=&-dt^2+\sum_{m=1}^5dx^mdx^m+H_5(dx^6dx^6+d{\bf y}\cdot d{\bf
y})\nonumber\\ e^{2\Phi}&=&H_5(y)=
{\alpha^\prime\over2\ell}\left({1\over|{\bf
y}|}+{1\over|{\bf y}-{\bf y}_0|}\right)\ ,
\end{eqnarray}
where $\bf y$ is a 3-vector in the $(x^7,x^8,x^9)$ plane. We have
placed one NS5-brane at ${\bf y}=0$ and the other at ${\bf
y}={\bf y}_0$. The condition~\cite{fivebranes,harvey}
$H_{mns}=\epsilon_{mns}^{\phantom{mns}r}\partial_r\Phi$ defines
$B_{6i}$ as a vector $\omega_i$ which satisfies $\nabla
H_5=\nabla{\times}{\bf\omega}$.  Applying the usual supergravity
$T$-duality rules gives:
\begin{eqnarray}
ds^2&=&-dt^2+\sum_{m=1}^5dx^mdx^m+H_5^{-1}(dx^6+\omega_i
dy^i)^2+H_5d{\bf y}\cdot d{\bf y}\ ,
\end{eqnarray}
which is the two-centre Gibbons--Hawking metric for the $A_1$ ALE
space, nonsingular because of the
$2\pi\alpha^\prime/\ell{\equiv}2\pi\ell^\prime$ periodicity of $x^6$,
and hyperK\"ahler because of the condition relating $H_5$ and $B_{6i}$
for a supersymmetric fivebrane solution.

There are four moduli associated with this solution, forming a
hypermultiplet in the $(5+1)$-dimensional theory. Three of them
constitute the vector ${\bf y}_0$ giving the separation between the
two centres. The fourth is a NS--NS 2-form flux,
$\Theta=\int_{\IP^1}B_2$, through the nontrivial two-cycle (a
$\IP^1$) in the space.  This $\IP^1$ is constructed as the locus of
$x^6$ circles along the straight line connecting the two centres ${\bf
y}=0$ and ${\bf y}={\bf y}_0$, where they shrink to zero size; it
has area $A=2\pi\ell^\prime|{\bf y}_0|$. These four numbers specify
the separation of the two NS5-branes in the $T_6$-dual picture.

In our case, we have ${\bf y}_0=0$, and so the branes are only
separated in $x^6$, corresponding to having shrunk the $\IP^1$ away.
The flux $\Theta$ is kept finite as we send $A$ to zero, and is the parameter
dual to $L$:
\begin{equation}\label{elldual}
\Theta = 2\pi \ell^\prime L \,.
\end{equation}
(The full NS5-branes solution, with dependence on $x^6$, can be
recovered in the duality by considering winding strings in the ALE
geometry~\cite{gregory}, doing a sum over those modes which is dual to
a Fourier transform of the fivebrane harmonic function on the $x^6$
circle.)

On the ALE space, the $W$-bosons for the enhanced $SU(2)$ gauge
symmetry of the $(5 + 1)$-dimensional theory are made from a D2-
and an anti D2-brane wrapped on the $\IP^1$, their masses being
$m=\mu_2g^{-1}\sqrt{(2\pi\ell^\prime)^2 {\bf y}_0\cdot{\bf
y}_0+\Theta^2}$, where the flux appears due to the
$\mu_2C_1{\wedge}B_2$ coupling; there is some induced D0-brane
charge.  Under $T_6$-duality, the stretched D$(p + 1)$-branes become
D$(p + 2)$-branes which are wrapped on the $\IP^1$, inducing some
D$p$-brane charge due to the $\mu_2 C_{p+1}{\wedge}B_2$ coupling.

So the configuration with $N$ D$(p + 1)$-branes stretched
between the two NS5-branes is dual to the same number of
D$(p + 2)$-branes wrapped on the two-cycle of an $A_1$ ALE space,
giving rise to $N$ effective D$p$-branes. The gauge coupling
in the resulting $(p+1)$-dimensional $SU(N)$ gauge theory
is given by:
\begin{equation}
g^2_{\rm YM}=
\Theta^{-1}g(2\pi)^p\alpha^{\prime(p-1)/2}\ .
\label{couples}
\end{equation}

As mentioned in the introduction, this configuration is $T$-dual to
$N$ D$(p~+~4)$ branes wrapped on a K3, of volume $V$, the parameter
dual to $\Theta$. The things we learned about this original
configuration translate into refinements of the dual pictures.  For
example, the nontrivial metric on moduli space corresponds to the
bending~\cite{edbend} of the NS5-branes away from being flat,
resulting from the D$(p+1)$-branes' pull on them (See figure
\ref{bendy}(b)).  Denoting the radial coordinate in the
$(x^{p+1},\ldots,x^5)$ directions as $r$, (as we did before), the
$x^6$ position of a NS5-brane is given by the equation $\nabla^2
x^6(r) = 0$, giving the smooth shape of the NS5-brane for large
enough $r$, {\it i.e.} far away enough from the details of the
junction itself.

The solution for the shape is $x^6(r) = \alpha +  \beta N
r^{p-3}$ ($p \neq 3$), or $\alpha N\log(r/\beta)$ (for $p = 3$), where
$\alpha$ and $\beta$ are constants set by $L$, the asymptotic
separation of the branes, and $g$ and $\alpha^\prime$. Using this, an
expression for the separation $L(r)$ of the NS5-branes is
\begin{eqnarray}
L(r)&=&L-{2\beta N\over r^{3-p}};\quad(p\neq3)\ ,\nonumber\\
L(r)&=&2\alpha N\log\left({r\over \beta}\right);\quad (p=3)\ .
\end{eqnarray}
Notice that this is precisely the functional behavior that we see in
the harmonic functions for the supergravity solution of the
D$p$-D$(p+4)$ system. The parameters $\alpha$ and $\beta$ can be fixed
completely by comparing to the large $r$ limit of probe moduli space
computation done there, although we will not do it here.

In the expression (\ref{coupling}) for the gauge
coupling, we should replace $L$ by our expression for $L(r)$, giving
the running of the coupling with position on the Coulomb branch.

We recover therefore the singularity in the Coulomb branch where the
gauge coupling diverges, when the separation $L(r)$ of the
NS5-branes is of order $\alpha^\prime$, resulting in an enhanced
gauge symmetry on the NS5-branes' (5+1)-dimensional
world-volume.  The enhan\c con is simply the $S^{4-p}$ of closest
approach of the NS5-branes in the $x^6$ direction.  Notice that
the probe brane we studied previously is a single stretched
D$(p + 1)$-brane moving in $(x^{p+1},\ldots,x^5)$ in this picture.

An $r$-dependence in the separation between the NS5-branes translates,
using (\ref{elldual}), into variable $W$-boson masses $m =
\mu_2g^{-1}\Theta(r)$ in the $(5+1)$-dimensional theory, giving the
enhan\c con locus where $\Theta(r)$ vanishes.  Using $\Theta(r)$ in
the formula (\ref{couples}) results in the divergence in the $(p +
1)$-dimensional gauge coupling at the enhan\c con locus.

%--------------------------------------------------------------------+
\newpage
\sect{Conclusions}

One notable result of this paper is a new mechanism that resolves a
large class of spacetime singularities in string theory.  This
involves a phenomenon, the resolution of a singularity by the
expansion of a system of branes in the transverse directions, which is
related to that which has recently arisen in other
forms~\cite{coul,noncom}.  One difference from \cite{coul} is that the
branes are found not at the singularity in the supergravity metric;
rather, the metric is modified by string/braney phenomena in the
manner that we have described.  Our result may point toward a more
general understanding of singularities in string theory.

In the gauge theory we have found a striking parallel between the
spacetime picture and the behavior of large-$N$ $SU(N)$ gauge
theories.  The most interesting open question is to find a weakly
coupled dual to the strongly coupled gauge theory; our results give
many hints in this direction.  There are a number of technical loose
ends, which include a more complete treatment of the D3-D7 case, and a
fuller understanding of the constraints of supersymmetry on the probe
moduli space.  Finally, there are many interesting generalizations,
including product gauge groups, the addition of hypermultiplets, and
rotation.

%--------------------------------------------------------------------+
\section*{Acknowledgments}
We would like to thank P. (Bruce Willis) Argyres, S. (Andrew Jackson)
Gubser, A. Hanany, A. Hashimoto, P. Ho{\u{r}}ava, A. Karch, R. Myers,
P. Pouliot, S. Sethi, A. Strominger, and C.  Vafa for helpful remarks
and discussions.  This work was supported in part by NSF grants
PHY94-07194, PHY97-22022, and CAREER grant PHY97-33173.

%--------------------------------------------------------------------+

%--------------------------------------------------------------------+
\end{document}